\documentclass{webofc}
\fancypagestyle{firstpage}{%
  \fancyhf{}%
  \renewcommand{\headrulewidth}{0pt}%
  \renewcommand{\footrulewidth}{0pt}%
  \fancyhead[R]{%
    \minipage[t][\baselineskip]{\linewidth}
      \small
      XXIII International Baldin Seminar on High Energy Physics Problems\\
       "Relativistic Nuclear Physics and Quantum Chromodynamics" \\
      to appear in EPJ Web of Conferences \\
    \endminipage
  }%
}
\usepackage[utf8]{inputenc}
\usepackage[varg]{txfonts}   

\wocname{}
\woctitle{}

\newcommand{\beq}{\begin{equation}}
\newcommand{\eeq}{\end{equation}}
\newcommand{\bea}{\begin{array}}
\newcommand{\eea}{\end{array}}
\newcommand{\beqa}{\begin{eqnarray}}
\newcommand{\eeqa}{\end{eqnarray}}

\def\beqa{\begin{eqnarray}}
\def\eeqa{\end{eqnarray}}

\begin{document}
\title{Study of lattice QCD at finite chemical potential using canonical ensemble approach}
%
%

\author{V.~G. Bornyakov\inst{1,3,4}\fnsep\thanks{\email{vitaly.bornyakov@ihep.ru}} \and
        D.~L. Boyda\inst{1,2,4} \and
        V.~A. Goy\inst{1,2,4} \and
        A.~V.~Molochkov\inst{1,4} \and
        Atsushi Nakamura\inst{1,5,6} \and
        A.~A. Nikolaev\inst{1,4} \and
        V.~I. Zakharov\inst{1,4}
}

\institute{School of Biomedicine, Far Eastern Federal University, 690950 Vladivostok, Russia
\and
           School of Natural Sciences, Far Eastern Federal University, 690950 Vladivostok, Russia
\and
           Institute for High Energy Physics NRC Kurchatov Institute, 142281 Protvino, Russia,
\and
           Institute of Theoretical and Experimental Physics NRC Kurchatov Institute, 117218 Moscow, Russia
\and
           Research Center for Nuclear Physics (RCNP), Osaka University, Ibaraki, Osaka, 567-0047, Japan,
\and
           Theoretical Research Division, Nishina Center, RIKEN, Wako 351-0198, Japan
}

\abstract{New approach to computation  of canonical partition functions in $N_f=2$  lattice QCD is presented.
  We compare results obtained by new method with results obtained by known method of hopping parameter expansion. We observe agreement between two methods indicating  validity of the new method. We use results for the
  number density obtained in the confining and deconfining phases at imaginary chemical potential to determine the phase transition line at real chemical potential.}
\maketitle
\section{Introduction}
\label{sec:introduction}

Recent results of heavy ion collision experiments at RHIC \cite{Adams:2005dq} and  LHC  \cite{Aamodt:2008zz}
shed some light on properties of the quark
gluon plasma and the position of the transition line in the temperature – baryon density plane.
New experiments will be carried out at FAIR (GSI) and NICA (JINR).
To fully
explore the phase diagram theoretically it is necessary to make computations at finite temperature and
finite baryon chemical potential. For finite temperature lattice QCD is the only ab-initio method available
and many results had been obtained. However, for finite baryon density lattice QCD faces the so-called
complex action problem (or sign problem).
Various proposals exist at the moment to solve this problem see, e.g. reviews~\cite{Muroya:2003qs,Philipsen:2005mj,deForcrand:2010ys} and yet it is still
very hard to get reliable results at $\mu_B/T>1$. Our work is devoted to developing the canonical partition function approach.

The fermion determinant at nonzero baryon chemical potential $\mu_B$, $\det\Delta(\mu_B)$, is in general not real.
This makes impossible to apply standard Monte Carlo techniques to computations with the partition function
\beq
Z_{GC}(\mu_q,T,V) = \int \mathcal{D}U (\det\Delta(\mu_q))^{N_f} e^{-S_G},
\label{Eq:PathIntegral}
\eeq
where $S_G$ is a gauge field action, $\mu_q=\mu_B/3$ is quark chemical potential,   $T=1/(aN_t)$ is temperature, $V=(aN_s)^3$ is volume, $a$ is lattice spacing, $N_t, N_s$ - number of lattice sites in time and space directions.

The canonical approach was studied in a number of papers \cite{deForcrand:2006ec,Ejiri:2008xt,Li:2010qf,Li:2011ee,Danzer:2012vw,Gattringer:2014hra,Fukuda:2015mva,Nakamura:2015jra}.
It is based on the following relations. Relation between
grand canonical partition function $Z_{GC}(\mu_q, T, V)$ and the canonical one $Z_C(n, T, V)$
called fugacity expansion:

\beq
Z_{GC}(\mu, T, V)=\sum_{n=-\infty}^\infty Z_C(n,T,V)\xi^n,
\label{ZG}	
\eeq
where
$\xi=e^{\mu_q/T}$ is the fugacity.
The inverse of this equation can be presented in the following form
\cite{Hasenfratz:1991ax}
\beq
Z_C\left(n,T,V\right)=\int_0^{2\pi}\frac{d\theta}{2\pi}
e^{-in\theta}Z_{GC}(\theta,T,V).
\label{Fourier}
\eeq
$Z_{GC}(\theta,T,V)$ is the grand canonical partition function
for imaginary chemical potential $\mu_q=i\mu_{qI} \equiv iT\theta$. Standard Monte Carlo simulations are
possible for this  partition function since fermionic determinant is real for imaginary $\mu_q$.

\begin{figure}[htb]
\centering
\includegraphics[width=0.35\textwidth,angle=0]{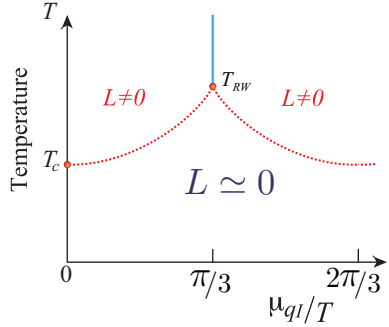}%
\vspace{0cm}
\caption{Schematical figure of Roberge-Weiss phase structure in the pure imaginary chemical
potential regions.}
\label{RW_ph_d}
\end{figure}

The QCD partition function  $Z_{GC}$ is a periodic function of $\theta$: $Z_{GC}(\theta) = Z_{GC}(\theta+2\pi/3)$.
This symmetry is called Roberge-Weiss symmetry \cite{Roberge:1986mm}. As a consequence of this periodicity
the canonical partition functions $Z_C(n,T,V)$  are nonzero only for $n=3k$.
 QCD possesses a rich phase structure at nonzero $\theta$, which depends on the number of
flavors $N_f$ and the quark mass $m$. This phase structure is shown in
Fig.~\ref{RW_ph_d}.  $T_c$ is the confinement/deconfinement crossover point at
zero chemical potential.  The line $(T \ge T_{RW},\mu_I/T=\pi/3)$ indicates the
first order phase transition.  On the curve between $T_c$
and $T_{RW}$, the transition is expected to change from the crossover to the first order
for small and large quark masses, see e.g. \cite{Bonati:2014kpa}.
 Quark number density $n_q$ for $N_f$ degenerate quark flavours is defined by the following equation:
\beq
\frac{n_{q}}{T^{3}} = \frac{1}{VT^{2}}\frac{\partial}{\partial \mu_q}\ln
Z_{GC}
=\frac{N_{f}N_{t}^{3}}{N_s^3 Z_{GC}} \int \mathcal{D}U e^{-S_G} (\det\Delta(\mu_q))^{N_f}
\mathrm{tr}\left[\Delta^{-1}\frac{\partial \Delta}{\partial \mu_q/T}\right].
\label{density1}
\eeq
It can be computed numerically for imaginary chemical potential. Note, that for the imaginary chemical potential $n_q$ is also purely imaginary: $n_q = i n_{qI}$.

From eqs.~(\ref{ZG}) and (\ref{density1}) it follows that densities $n_{q}$ and $n_{qI}$ are related  to $Z_C(n,T,V)$ (below we will use the notation $Z_n$ for the ratio
$Z_C(n,T,V) / Z_C(0,T,V)$)  by equations
\beq
\label{density2}
n_{q}/T^3  = {\cal{N}}\frac{2\sum_{n>0} n Z_n \sinh(n\theta)}{1+2\sum_{n>0} Z_n \cosh(n\theta)},\,\,
n_{qI}/T^3  = {\cal{N}}\frac{2\sum_{n>0} n Z_n \sin(n\theta)}{1+2\sum_{n>0} Z_n \cos(n\theta)}\,,
\eeq
where ${\cal{N}}$ is a normalization constant, ${\cal{N}}=\frac{N_t^3 }{N_s^3}$.
Our suggestion is to compute $Z_n$ using equation (\ref{density2}) for $n_{qI}$.

One can compute $Z_{GC}(\theta,T,V)$ using numerical data for $n_{qI}/T^3$ via numerical integration
\beq
L_Z(\theta) \equiv \log\frac{Z_{GC}(\theta,T,V)}{Z_{GC}(0,T,V)}  = - V \int_{0}^{\theta} d \tilde{\theta}~n_{qI}(\tilde{\theta})\,,
\label{integration_1}
\eeq
where we omitted $T$ and $V$ from the grand canonical partition function
notation.
Then $Z_n$ can be computed as
\beq	
Z_n = \frac{\int_0^{2\pi}\frac{d\theta}{2\pi} e^{-in\theta} e^{L_Z(\theta)} }{ \int_0^{2\pi}\frac{d\theta}{2\pi}
 e^{L_Z(\theta)} }
\label{Fourier_2}
\eeq

In our work we use modified version of this approach \cite{Bornyakov:2016}.
Instead of numerical integration in (\ref{integration_1})
we fitted $n_{qI}/T^3$ to theoretically motivated functions
of $\mu_{qI}$. It is known that the density of noninteracting  quark gas is described  by
\beq
n_q/T^3 = N_f \Bigl ( 2\frac{\mu_q}{T}  + \frac{2}{\pi^2} \Bigl (\frac{\mu_q}{T} \Bigr )^3 \Bigr ).
\eeq
We thus fit the data for $n_{qI}$ to an odd power polynomial of $\theta$
\beq
n_{qI}(\theta)/T^3 = \sum_{n=1}^{n_{max}} a_{2n-1} \theta^{2n-1}\,,
\label{eq_fit_polyn}
\eeq
in the deconfining phase. This type of the fit was also used in Ref.~\cite{Takahashi:2014rta} and Ref.~\cite{Gunther:2016vcp}.

In the confining phase (below $T_c$) the hadron resonance gas model provides
good description of the chemical potential dependence of thermodynamic observables
\cite{Karsch:2003zq}.
Thus it is reasonable to fit the density to a Fourier expansion
\beq
n_{qI}(\theta)/T^3 = \sum_{n=1}^{n_{max}} f_{3n} \sin(3n \theta)
\label{eq_fit_fourier}
\eeq
Again this type of the fit was  used in Ref.~\cite{Takahashi:2014rta} and conclusion was made that it works well.

To demonstrate our method we made simulations of the lattice QCD with $N_f=2$ clover improved Wilson quarks and Iwasaki improved gauge field action, for detailed definition of the lattice action see \cite{Bornyakov:2016}.
We simulate $16^3 \times 4$ lattices at temperatures $T/T_c=1.35, 1.20$ and 1.08 in the deconfinemnt phase  and $0.99, 0.93, 0.84$ in the confinement phase along the line of constant physics with
 $m_{\pi}/m_{\rho}=0.8$. All parameters of the action, including $c_{SW}$ value were borrowed from the WHOT-QCD collaboration paper~\cite{Ejiri:2009hq}. We compute the number density on samples of $N_{conf}$ configurations with $N_{conf}=1800$, using every 10-th trajectory produced with Hybrid Monte Carlo algorithm.
\section{Results}
\label{mainresults}

\begin{figure}[htb]
\begin{center}
	\begin{minipage}[t]{0.49\textwidth}
    	\includegraphics[width=0.73\textwidth,angle=-90]{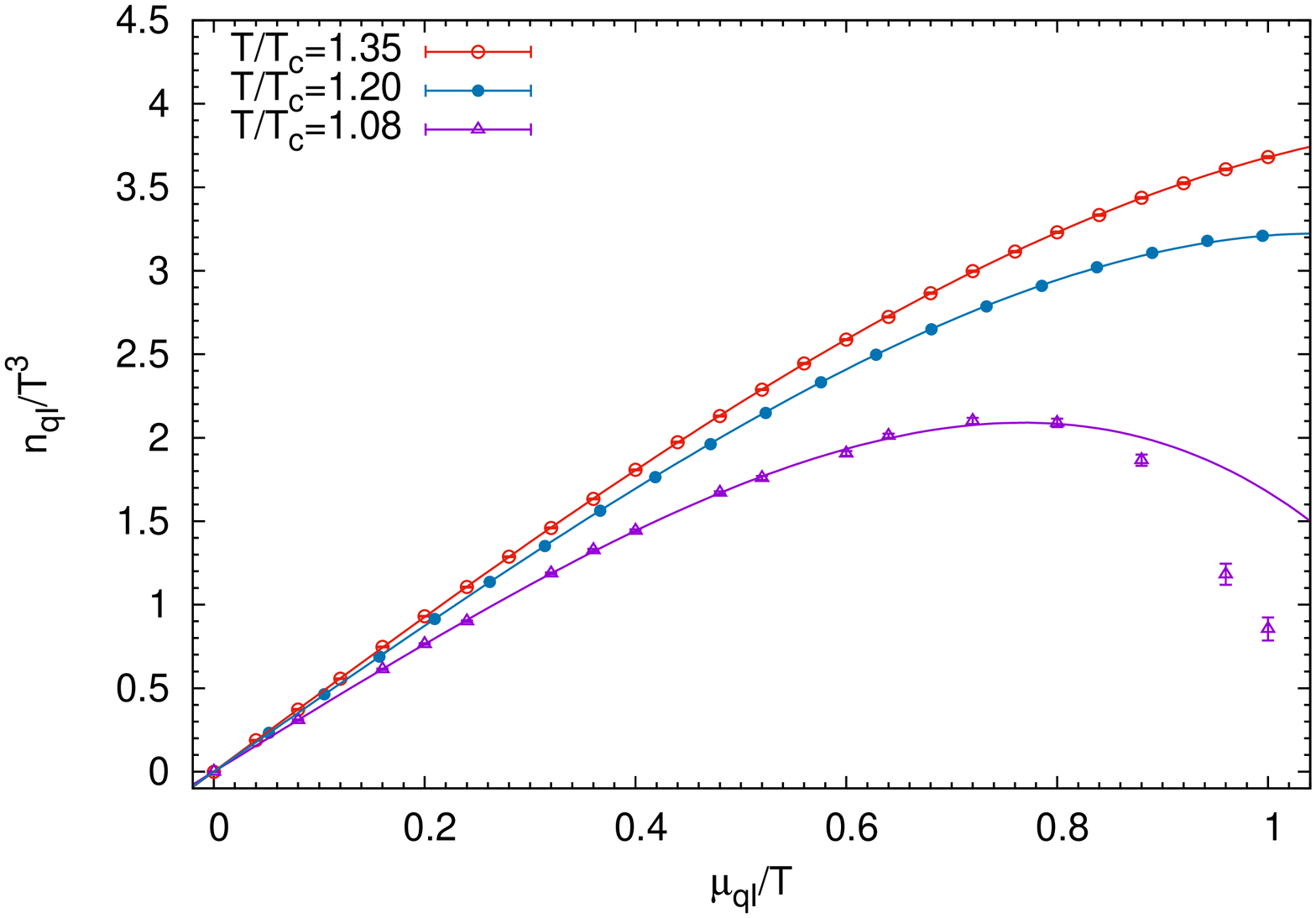}
    \caption{Imaginary density as function of $\theta$ in the deconfinement phase at temperatures $T/T_c=1.35, 1.20, 1.08$.
The curves show fits to function (\ref{eq_fit_polyn}). }
\label{density_deconf}
    \end{minipage}
\hfill
    \begin{minipage}[t]{0.49\textwidth}
    \vspace{-0mm}
	\includegraphics[width=0.77\textwidth,angle=-90]{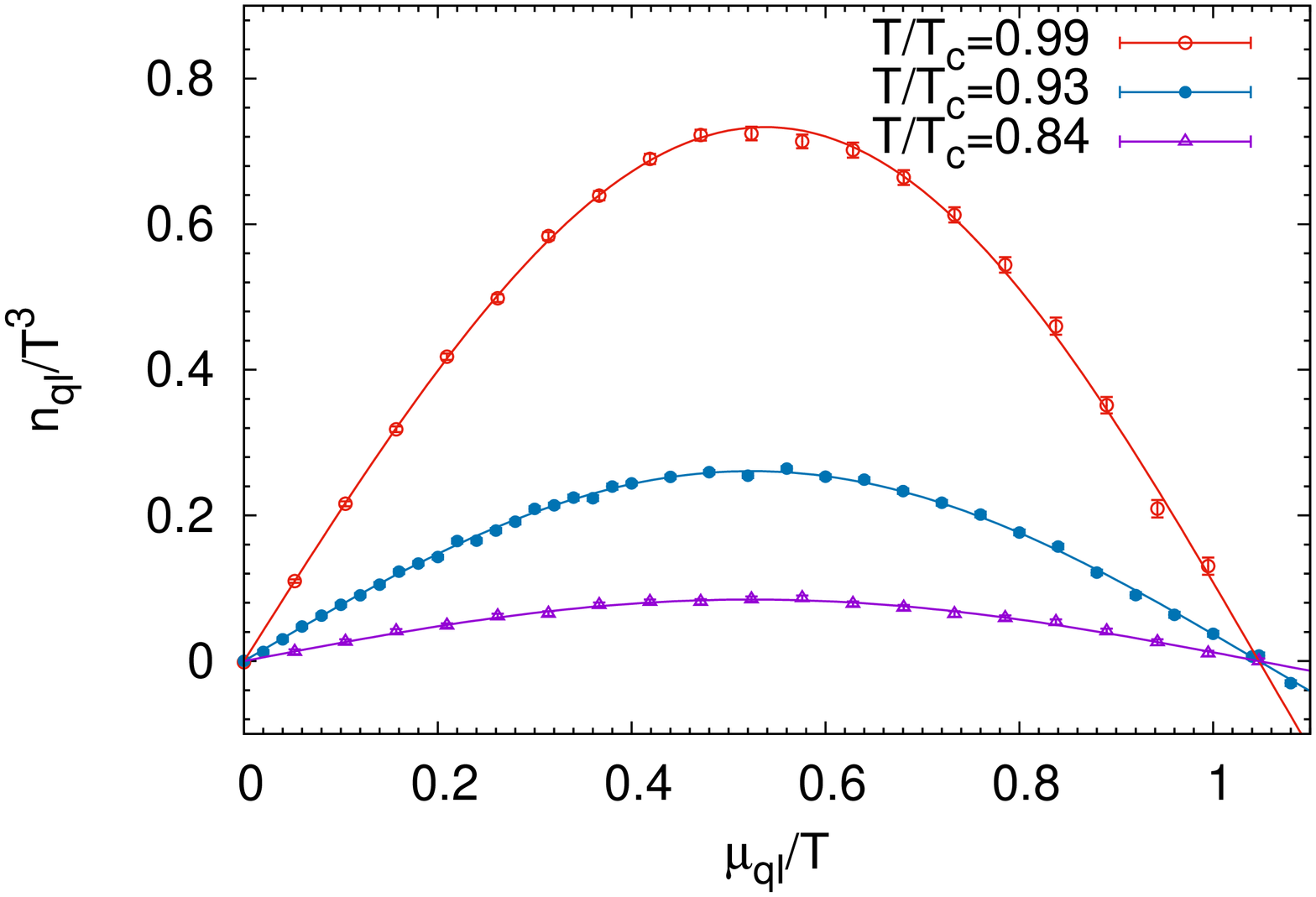}
    \vspace{-4mm}
	\caption{Imaginary density as function of $\theta$ for temperatures in the confining phase. The curves show fits to (\ref{eq_fit_fourier}) with $n_{max} = 1$ for $T/T_c = 0.84, 0.93$ and $ n_{max}=2$ for $T/T_c = 0.99$.}
\label{density_conf}
    \end{minipage}
\end{center}
\end{figure}
In Fig.~\ref{density_deconf} and Fig.~\ref{density_conf}  we show numerical results for $n_{qI}$ as a function of $\theta$ in
deconfining and confining phases, respectively \cite{Bornyakov:2016}.

Note, that behavior of  $n_{qI}$  at $T=1.08$ is different from that at higher temperatures. This temperature is below $T_{RW}$ and at  $\theta=\pi/3$ there is no first order
phase transition, $n_{qI}$ is continuous. Instead there is a crossover to the confinement phase
at about $\theta=0.92(2)$.

It is not yet clear how to fit the data over the range of $\mu_{qI}$ covering both deconfining and confining phase.
Here we use fit to  function (\ref{eq_fit_polyn}) with $n_{max}=3$  over the range $[0,0.8]$, i.e. including only the deconfining phase.
In this case we should consider the fit as a Taylor expansion.

\begin{figure}[htb]
\begin{center}
	\begin{minipage}[t]{0.49\textwidth}
    	\includegraphics[width=0.73\textwidth,angle=-90]{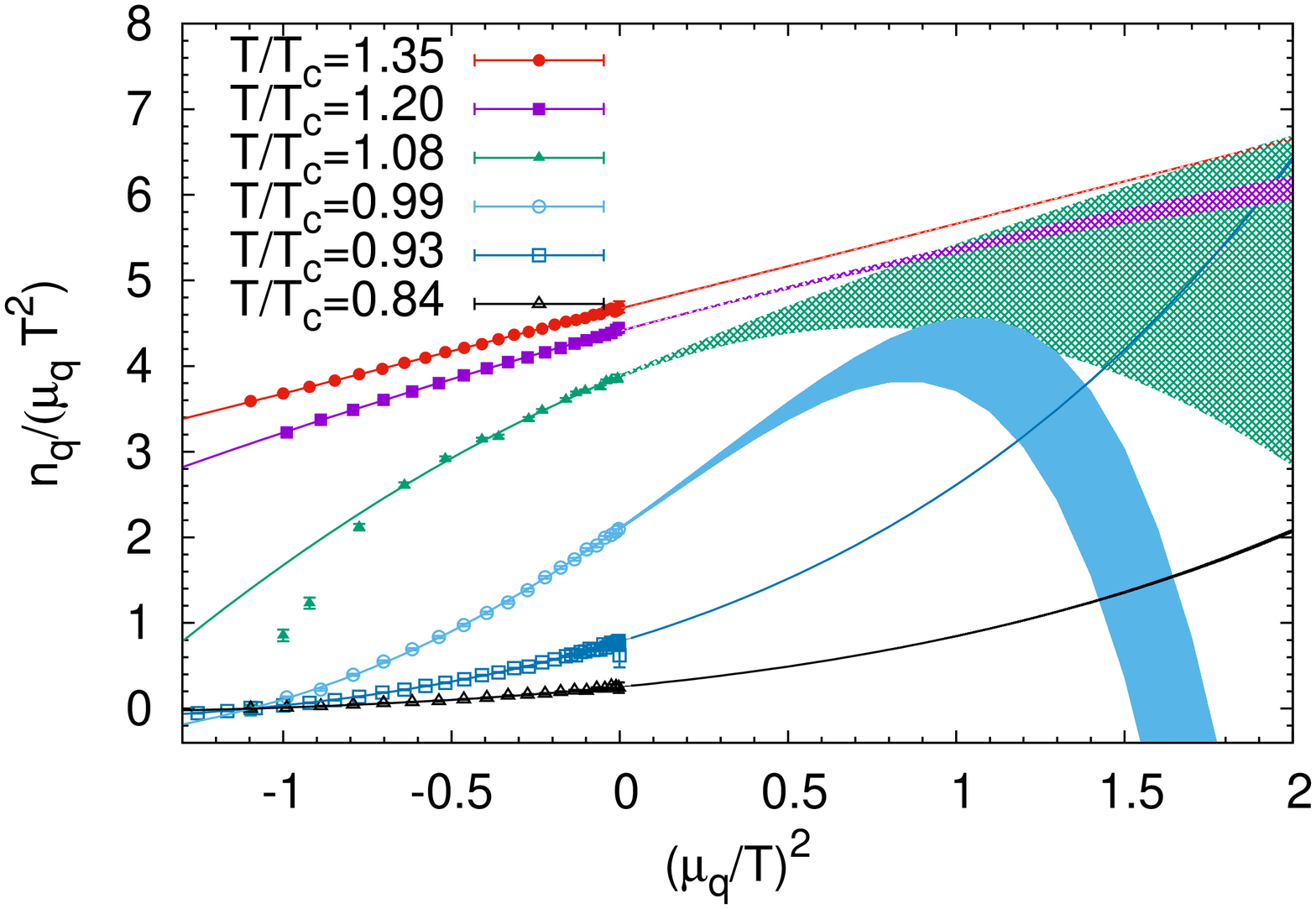}
    \caption{Analytical continuation for the number density vs. $\mu_q^2$.
The curves show respective fits eq.~(\ref{eq_fit_polyn}) or eq.~(\ref{eq_fit_fourier}). The width of the curves indicates the statistical error of extrapolation to $\mu_q^2 > 0$.}
\label{dens_anal_cont}
    \end{minipage}
\hfill
    \begin{minipage}[t]{0.49\textwidth}
    \vspace{-0mm}
	\includegraphics[width=0.77\textwidth,angle=-90]{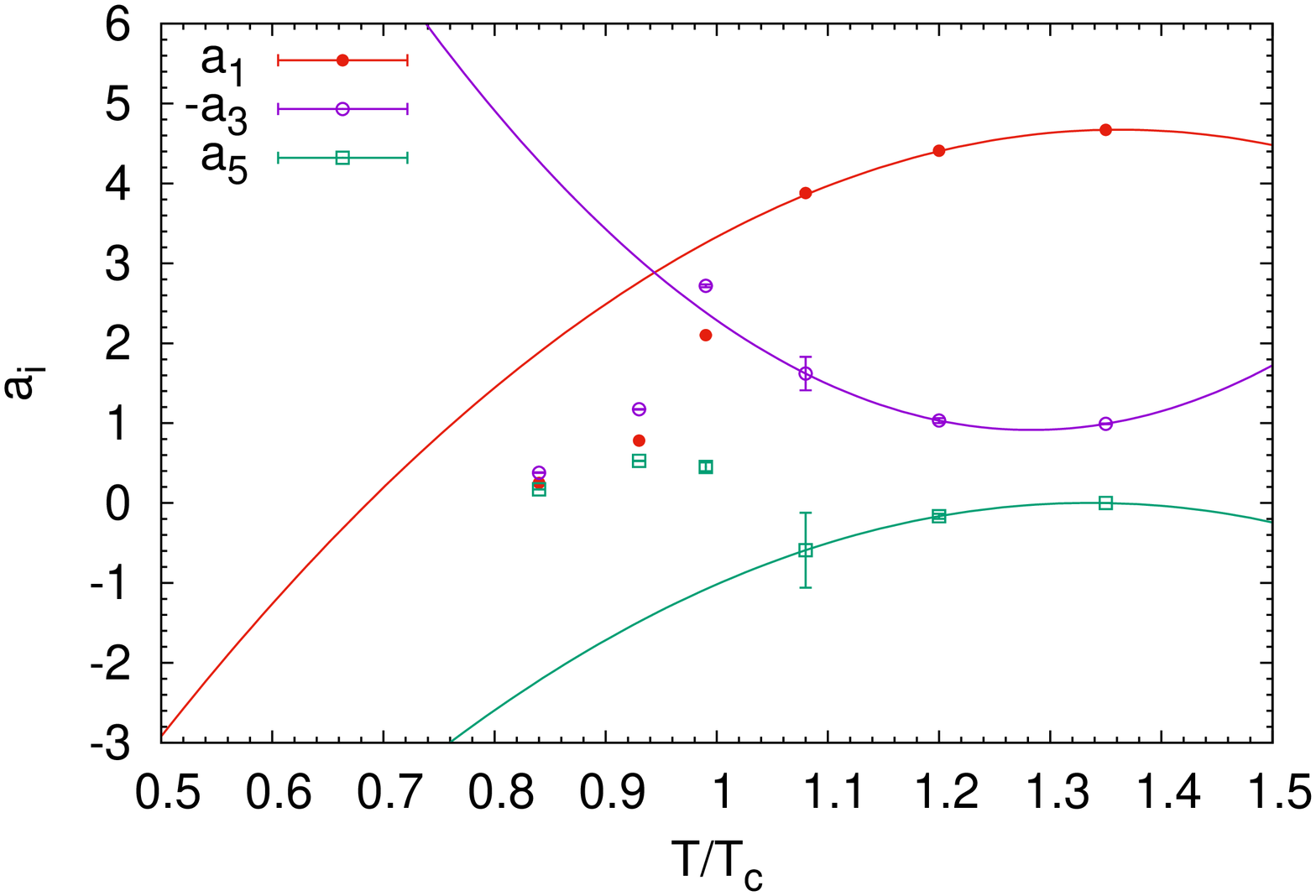}
    \vspace{-4mm}
	\caption{Constants $a_i$ of eq.~(\ref{eq_fit_polyn}) for six values of temperature. The curves show polynomial fits over $T>T_c$ range.}
\label{constants}
    \end{minipage}
\end{center}
\end{figure}

We  computed $Z_n$ using new procedure described in the previous section  and compared with results obtained with use of
the hopping parameter expansion. We found good agreement between two methods indicating that the new method works well \cite{Bornyakov:2016}.
This allows us to make analytical continuation to the real values of $\mu_q$ beyond the Taylor expansion validity range (for all temperatures
apart from $T/T_c=1.08$). Respective results are shown in Fig.~\ref{dens_anal_cont}.
Using the results for the number density $n_q$ we can compute the temperature of the transition from the hadron  phase to quark-gluon plasma phase using the following procedure.
Our results for the number density at temperatures $T>T_c$ as function of the chemical potential $\mu_q$
are reliable even for large $\mu_q$ for $T>T_{RW}$, while for $T_c < T < T_{RW}$ they are reliable at the moment for small $\mu_q$
only. We can use these results to compute the pressure $\Delta P_{deconf}(T,\mu_q) = P(T,\mu_q) - P(T,0)$ as function of $\mu_q$ and then extrapolate pressure for fixed $\mu_q$ to
temperatures $T<T_c$.  To get good extrapolation we need pressure computed for more values of temperature than is available now
but three values which we have in this work is enough to demonstrate the idea. We then find the transition temperature $T_c(\mu_q)$
solving numerically equation $\Delta P_{deconf}(T,\mu_q) = \Delta P_{conf}(T,\mu_q)$,
where $\Delta P_{conf}(T,\mu_q)$ is pressure computed from results for the number density $n_q$ we obtained in the confinement phase.
In this paper we use extrapolation for the coefficients in (\ref{eq_fit_polyn}) rather than for pressure itself.
This extrapolation is shown in Fig.~\ref{constants}. We fitted the data for $a_i, i=1,3,5$ by a polynomial $b_0+b_1\frac{T}{T_c}+b_2\left(\frac{T}{T_c}\right)^2$ and
then computed the extrapolated values of these parameters at $T/T_c=0.93$ and 0.84. The extrapolated values were used to compute
$\Delta P_{deconf}(T,\mu_q)/T^4$ as
\beq
\frac{\Delta P_{deconf}(T,\mu_q)}{T^4} = \frac{a_1}{2} \left(\frac{T}{T_c}\right)^2 - \frac{a_3}{4} \left(\frac{T}{T_c}\right)^4 + \frac{a_5}{6} \left(\frac{T}{T_c}\right)^6
\label{eq_pressure_deconf}
\eeq
\begin{figure}[htb]
\begin{center}
	\begin{minipage}[t]{0.49\textwidth}
    	\includegraphics[width=0.73\textwidth,angle=-90]{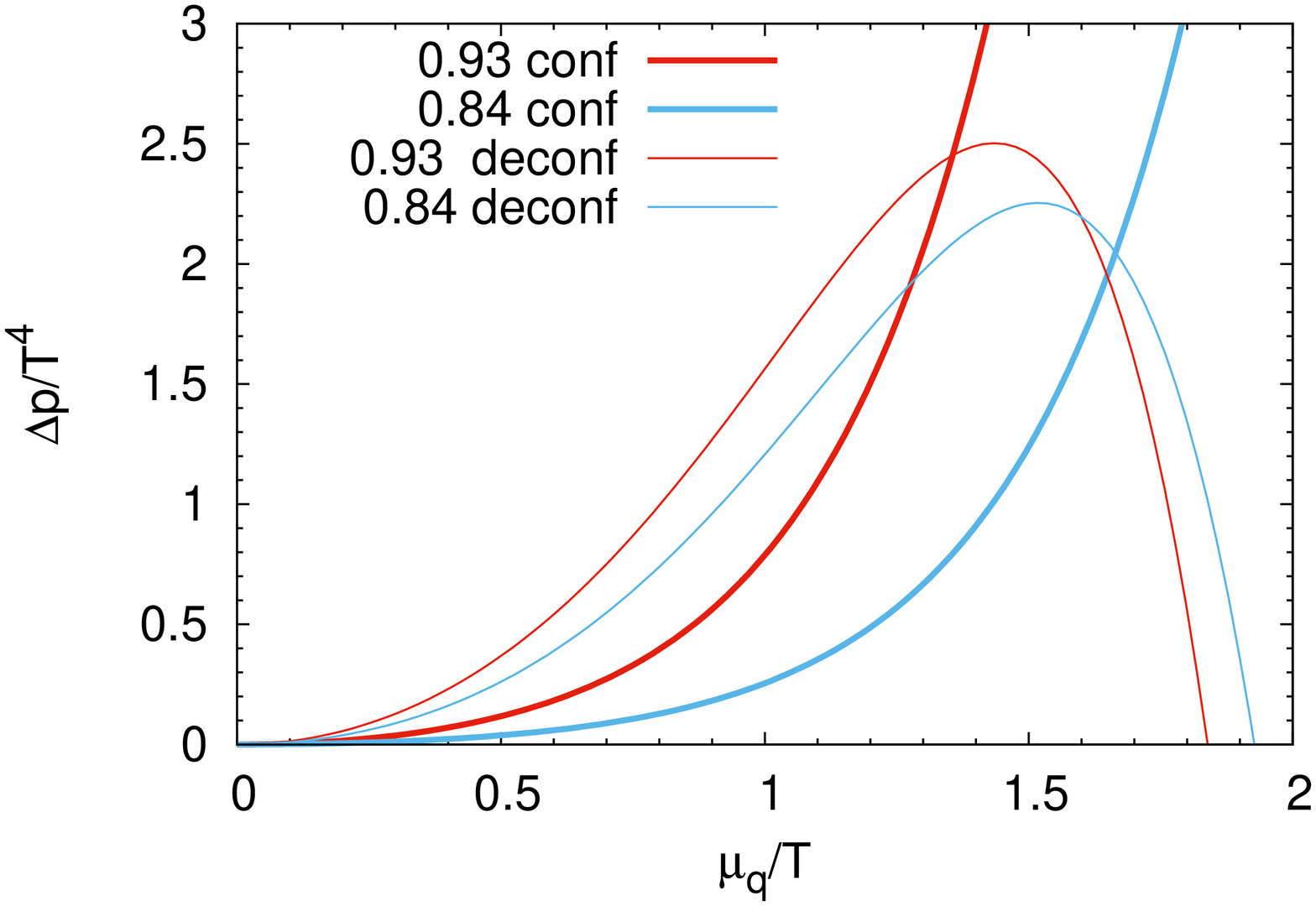}
    \caption{Pressure computed via eq.~(\ref{eq_pressure_deconf}) (denoted by 'deconf', thin curves) and by integration of eq.~(\ref{eq_fit_fourier})
    (denoted by 'conf', thick curves) for $T/T_c=0.93, 0.84$.
}
\label{fig_pressure}
    \end{minipage}
\hfill
    \begin{minipage}[t]{0.49\textwidth}
    \vspace{-0mm}
	\includegraphics[width=0.77\textwidth,angle=-90]{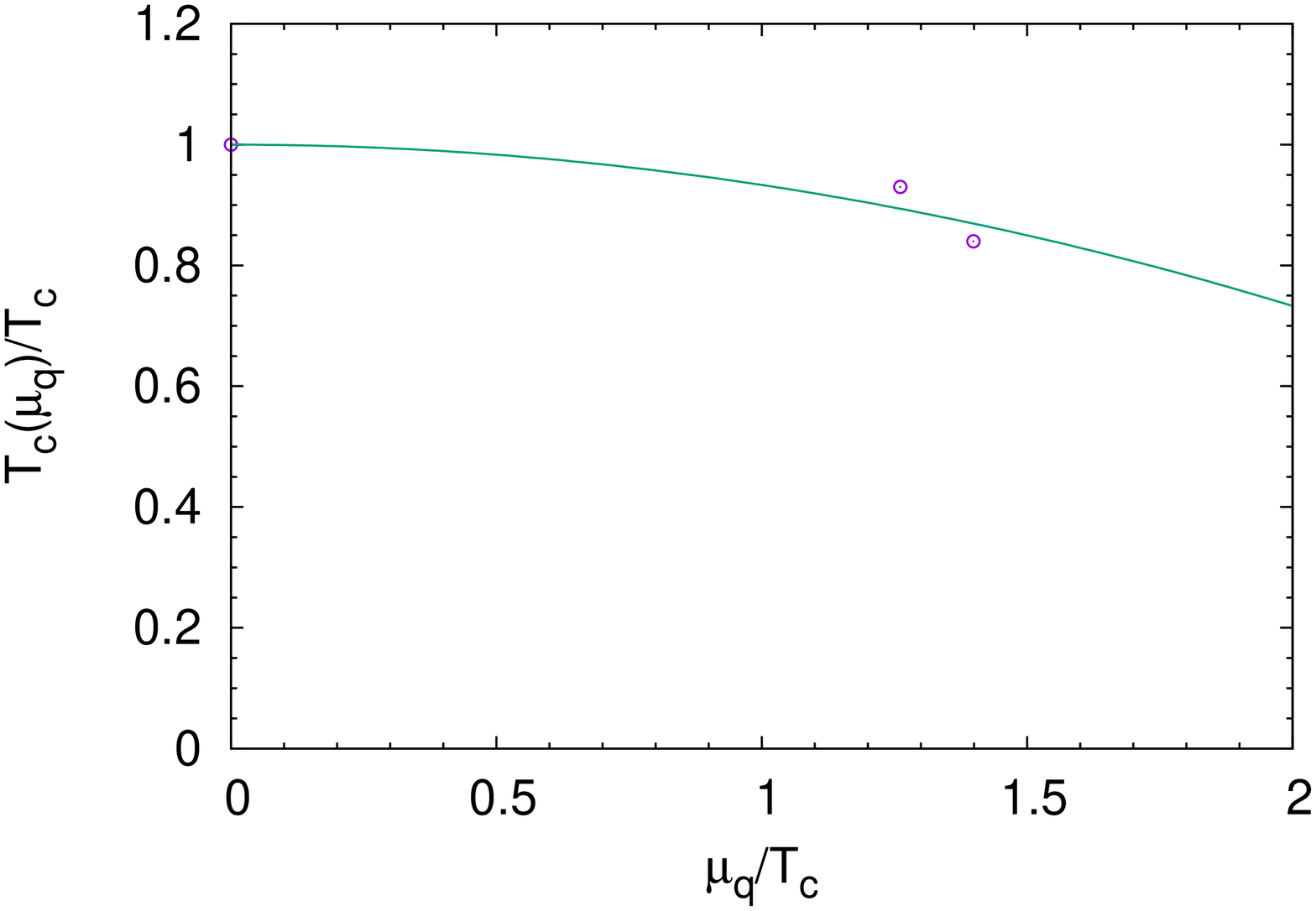}
    \vspace{-4mm}
	\caption{Transition line in the temperature - chemical potential plane. The curve show quadratic fit.}
\label{fig_Tc}
    \end{minipage}
\end{center}
\end{figure}
We show results in Fig.~\ref{fig_pressure} together with pressure computed by integration of eq.~(\ref{eq_fit_fourier}) for $T/T_c=0.93, 0.84$. The crossing points determine values of $\mu_q$ on transition line at respective value of temperature.
We show the coordinates of these crossing points in  Fig.~\ref{fig_Tc} together with fit of the form
\beq
T_c(\mu_q)/T_c = 1 - C \left(\mu_q/T_c \right)^2
\eeq
The result for the fit parameter is  $C=0.07$.
We have to emphasize that in Figs.~\ref{fig_pressure}  and ~\ref{fig_Tc} we do not show statistical or systematic errors
since we are not yet able to compute them. These figures are shown to illustrate our idea about the method to compute
the transition temperature dependence on the chemical potential. Still it is encouraging that the value of the coefficient
$C$ obtained is of correct order of magnitude. We can compare with Refs.~\cite{deForcrand:2002hgr,Wu:2006su} where $C=0.051(3)$ and  $C=0.065(7)$ respectively
were found in lattice QCD with $N_f=2$. In both papers analytical continuation of $\frac{T_c(\mu_q)}{T_c}$ from imaginary $\mu_q$
was used.

Thus we  presented new method to compute the canonical partition functions $Z_n$.
It is based on fitting of the imaginary number density for all values of imaginary
chemical potential to the theoretically motivated
fitting functions:  polynomial fit (\ref{eq_fit_polyn}) in the deconfinement
phase for $T$ above $T_{RW}$ and Fourier-type fit (\ref{eq_fit_fourier}) in the confinement phase.
We also explained how the transition line at real $\mu_q$ can be computed using results we obtained for the number density $n_q$.
We are planning to increase the number of temperature values to improve the precision of this method.
It is worth to note that the method of direct analytical continuation for $\frac{T_c(\mu_q)}{T_c}$  from
imaginary chemical potential can be applied to our data. This will help us to improve the method presented here.

\noindent
{\bf Acknowledgments} \\
This work was completed due to support by
RSF grant under contract 15-12-20008. Computer simulations were performed on the FEFU GPU cluster Vostok-1 and  MSU 'Lomonosov' supercomputer.


\begin{thebibliography}{23}

\bibitem{Adams:2005dq}
J.~Adams et~al. (STAR), Nucl. Phys. \textbf{A757}, 102 (2005),
  \texttt{nucl-ex/0501009}

\bibitem{Aamodt:2008zz}
K.~Aamodt et~al. (ALICE), JINST \textbf{3}, S08002 (2008)

\bibitem{Muroya:2003qs}
S.~Muroya, A.~Nakamura, C.~Nonaka, T.~Takaishi, Prog. Theor. Phys.
  \textbf{110}, 615 (2003), \texttt{hep-lat/0306031}

\bibitem{Philipsen:2005mj}
O.~Philipsen, PoS \textbf{LAT2005}, 016 (2006), [PoSJHW2005,012(2006)],
  \texttt{hep-lat/0510077}

\bibitem{deForcrand:2010ys}
P.~de~Forcrand, PoS \textbf{LAT2009}, 010 (2009), \texttt{1005.0539}

\bibitem{deForcrand:2006ec}
P.~de~Forcrand, S.~Kratochvila, Nucl. Phys. Proc. Suppl. \textbf{153}, 62
  (2006), [,62(2006)], \texttt{hep-lat/0602024}

\bibitem{Ejiri:2008xt}
S.~Ejiri, Phys. Rev. \textbf{D78}, 074507 (2008), \texttt{0804.3227}

\bibitem{Li:2010qf}
A.~Li, A.~Alexandru, K.F. Liu, X.~Meng, Phys. Rev. \textbf{D82}, 054502 (2010),
  \texttt{1005.4158}

\bibitem{Li:2011ee}
A.~Li, A.~Alexandru, K.F. Liu, Phys. Rev. \textbf{D84}, 071503 (2011),
  \texttt{1103.3045}

\bibitem{Danzer:2012vw}
J.~Danzer, C.~Gattringer, Phys. Rev. \textbf{D86}, 014502 (2012),
  \texttt{1204.1020}

\bibitem{Gattringer:2014hra}
C.~Gattringer, H.P. Schadler, Phys. Rev. \textbf{D91}, 074511 (2015),
  \texttt{1411.5133}

\bibitem{Fukuda:2015mva}
R.~Fukuda, A.~Nakamura, S.~Oka, Phys. Rev. \textbf{D93}, 094508 (2016),
  \texttt{1504.06351}

\bibitem{Nakamura:2015jra}
A.~Nakamura, S.~Oka, Y.~Taniguchi, JHEP \textbf{02}, 054 (2016),
  \texttt{1504.04471}

\bibitem{Hasenfratz:1991ax}
A.~Hasenfratz, D.~Toussaint, Nucl. Phys. \textbf{B371}, 539 (1992)

\bibitem{Roberge:1986mm}
A.~Roberge, N.~Weiss, Nucl. Phys. \textbf{B275}, 734 (1986)

\bibitem{Bonati:2014kpa}
C.~Bonati, P.~de~Forcrand, M.~D'Elia, O.~Philipsen, F.~Sanfilippo, Phys. Rev.
  \textbf{D90}, 074030 (2014), \texttt{1408.5086}

\bibitem{Bornyakov:2016}
V.G. Bornyakov, D.L. Boyda, V.A. Goy, A.V. Molochkov, A.~Nakamura, A.A.
  Nikolaev, V.I. Zakharov (2016), \texttt{1611.04229}

\bibitem{Takahashi:2014rta}
J.~Takahashi, H.~Kouno, M.~Yahiro, Phys. Rev. \textbf{D91}, 014501 (2015),
  \texttt{1410.7518}

\bibitem{Gunther:2016vcp}
J.~Gunther, R.~Bellwied, S.~Borsanyi, Z.~Fodor, S.D. Katz, A.~Pasztor, C.~Ratti
  (2016), \texttt{1607.02493}

\bibitem{Karsch:2003zq}
F.~Karsch, K.~Redlich, A.~Tawfik, Phys. Lett. \textbf{B571}, 67 (2003),
  \texttt{hep-ph/0306208}

\bibitem{Ejiri:2009hq}
S.~Ejiri, Y.~Maezawa, N.~Ukita, S.~Aoki, T.~Hatsuda, N.~Ishii, K.~Kanaya,
  T.~Umeda (WHOT-QCD), Phys. Rev. \textbf{D82}, 014508 (2010),
  \texttt{0909.2121}

\bibitem{deForcrand:2002hgr}
P.~de~Forcrand, O.~Philipsen, Nucl. Phys. \textbf{B642}, 290 (2002),
  \texttt{hep-lat/0205016}

\bibitem{Wu:2006su}
L.K. Wu, X.Q. Luo, H.S. Chen, Phys. Rev. \textbf{D76}, 034505 (2007),
  \texttt{hep-lat/0611035}

\end{thebibliography}

\end{document}